\documentclass[english, 12pt]{article}

\RequirePackage[OT1]{fontenc}
\RequirePackage{amsthm,amsmath, amssymb}
\RequirePackage[round, longnamesfirst]{natbib}
\RequirePackage[colorlinks, linkcolor=blue, anchorcolor=blue, citecolor=blue,
filecolor=blue, urlcolor=blue]{hyperref}
\usepackage{bbm}
\usepackage{subfigure}
\usepackage{graphicx}
\usepackage[english]{babel}
\usepackage{setspace}
\usepackage[margin=1.2in]{geometry}

\graphicspath{{figures/}}

\title{Comparing dominance of tennis' big three via multiple-output Bayesian quantile regression models}
\author{Bruno Santos\footnote{e-mail: b.santos@kent.ac.uk}}
\date{University of Kent}

\begin{document}
\maketitle

\onehalfspacing

\begin{abstract}
Tennis has seen a myriad of great male tennis players throughout its history and we are often interested in the discussion of who is/was the greatest player of all time. While we do not try to answer this question here, we delve into comparing some key statistics related to dominance over their opponents for the male players with the most Grand Slam titles, currently: Djokovic, Federer and Nadal, in alphabetical order. Here we consider the minutes played and the relative points in each of their completed matches, as a measure of dominance against other players. We consider important covariates such as surface, win or loss, type of tournament and whether their opponent was a top 20 ranked player in the world or not, to create a more complete comparison of their performance. We consider a Bayesian quantile regression model for multiple-output response variables to take into account the dependence between minutes and relative points won. This approach is compelling since we do not need to choose a probability distribution for the joint probability distribution of our response variable. Our results agree with the common intuition of Nadal's superiority in clay courts, Federer's superiority in grass courts and Djokovic's superiority in hard courts given their success in each of these surfaces; though Nadal's dominance in clay court games is unique. Federer shows his dominance regarding minutes spent in the court in wins, while Djokovic takes the edge when considering the dimension of relative points won, for most of the comparisons. While minutes can be directly connected to style of play, the relative points dimension could express more directly different levels of advantage over their opponent, in which Djokovic seems to be the overall leader in this analysis.
\end{abstract}

\section{Introduction}

Between the Wimbledon tournament in 2003 and the US Open in 2021, there were 74 Grand Slam tournaments played in competitive tennis. These tournaments are the most prestigious and the most important tournaments in tennis, given their points and money rewarded to players. In the men's side, in 60 of those 74 tournaments the winner of the tournament was in a small group of three very special players, also known in some media outlets as the "Big Three": Novak Djokovic, Roger Federer and Rafael Nadal. This is a quite long period of domination in the sport, where these three great players won more than 80\% of their matches and have amassed almost more than 3000 wins, with Djokovic just the one short of 1000 careers wins, with 978 at the time of writing this manuscript. This is certainly a measure of dominance of these players against all the others in the Association of Tennis Professionals (ATP) tour. Given this dominance over the years, an interesting question could be posed as to find out who is the more dominant among them.

There is certainly a great deal of debate if either one of these great three can be considered as the Greatest of All Time (GOAT) in the men's side. In fact, \citet{baker:14} studied the games of the Open Era and concluded that Roger Federer was a good choice for the GOAT title with all the available information regarding games played up until 2014. This is definitely an interesting discussion, but when one tries to answer this kind of question, it needs to use information or variables available for all players. In this case, one possible approach is try to estimate the probability of winning each match given the opponent. This could give an estimate of strength for each player, which could then be compared across different periods of tennis. The problem with this type of method is that it disregards possible differences between players from different eras, as far as conditioning, training regimes, ease (or difficulty) of travelling, level of play across the ATP tour, among others. The evolution of the game itself, for instance, regarding the material being used in the rackets could certainly have a difference in how these players have performed over time. For this reason, we avoid the comparison between players of different eras. 

Other modelling efforts with tennis data have looked at different questions as well. \citet{knottenbelt:12} considered a hierarchical Markov model to explain the probability of a player winning a match. Still on estimating this probability of winning a match, \citet{mchale:11} examined a Bradley-Terry type of model, taking into account the surfaces as well to forecast the final result; \citet{klaassen:03} proposed a method to update the probabilities of winning during the match as well and not just at the beginning. \citet{delCorral:10} used the difference of rankings to predict the outcomes of matches in Grand Slams. \citet{gorgi:19} considered a high dimensional dynamic model with player and surface specific parameters that was able to give better forecasts and it could be used to rank players as well. 

Here in this paper we focus our attention to data only pertaining to these three great players, who currently lead the grand slam titles list and are tied with 20 wins apiece: Djokovic, Federer and Nadal. The reason for not to include data from different players is the assumption that due to their extraordinary career path, their variation might not be explained by standards of regular players in the tour. For example, the model proposed by \citet{gorgi:19} predicts that Nadal had a 60\% probability of winning a match against Federer in a clay court game in February of 2017, which sounds slightly conservative given their record in clay courts games with 13 wins to Nadal and only 2 to Federer at that point. This is not an attempt of dismissing their work, but rather an indication that using other players information to get a possible baseline estimate and adding player specific components to explain these player abilities might not be appropriate for the big three.  

The variables considered in this analysis were chosen to approximate the idea of dominance or superiority between players during a tennis match. While one can agree that a quick match is not equivalent to being an "easy" win for one of the sides or a dominant win, we often observe that an easier match for a specific player will take less time on the clock in comparison with a tighter match. Therefore, one can think of duration of the match as a necessary condition, but not a sufficient one to approximate dominance of a particular match. Also, when one considers each single point separately, we could think that every player will try to win every point, even though sometimes their effort might not be the same at all points. How many points a player wins against another one relatively could give an idea of superiority in this sense. For instance, \citet{klaassen:01} posed the question whether points in tennis are independent and identically distributed and found out that the weaker a player is the stronger would be the effect contrary to their question. Even though, winning more points is not necessary a condition for winning the match, we believe that the relative points won does approximate the notion of dominance within a match. For example, we could use points won divided by points lost in a given match to measure this effect. With the aim to compare the dominance of these players against their opponents, we consider then minutes of each match for these players and relative points won as our response variable. Assuming that these variables are correlated, modelling them jointly certainly will bring more information.

In our modelling analysis, we consider Bayesian quantile regression models for multiple output response variables, in order to consider the information in both variables in the discussion of our results. The reason for this choice is the possibility of modelling these variables jointly, therefore using their correlation to build a more complete picture of their variation, but without assuming any probability distribution. Quantile regression models have been used for a number of applications since their inception by \citet{koenker:78} and have shown a great value to explain conditional quantiles, with flexibility to approximate these measures even without assuming any probability distribution for the response variable. A multivariate proposal was discussed by \citet{hallin:10}, where a directional approach is taken to estimate each quantile and a connection between quantile regions and depth regions was made. We consider the results of \citet{santos:20}, where the authors considered the problem of estimating several quantiles of interest and proposed a method to obtain noncrossing quantiles in this multivariate setting. Nevertheless, here we focus in just one quantile, $\tau = 0.25$, in order to have a more central quantile region to compare the performance of these players given the different covariates. Also, in our analysis the results were rather similar considering other values for $\tau$.

We organize this manuscript in the following manner. In the next section, we present the data set and its details regarding filters used to select the observations considered in the modelling part. In Section~\ref{sec_theory}, we give a brief introduction to Bayesian quantile regression models for multiple-output variables, as proposed by \citet{santos:20}. In the following section, we present our main results, where we compare the outputs of the model for different combinations of the covariates. We finish this manuscript with our final comments on Section~\ref{concluding}.

\section{Data set} \label{data_section}

We consider the data set organized by Jeff Sackman and available at the following repository: \url{https://github.com/JeffSackmann/tennis_atp}. We use all matches between 1998, the year where Federer started playing professionally, and until the US Open of 2020. For our analysis, we are interested only in the matches, where one of the players is either Roger Federer, Novak Djokovic or Rafael Nadal, so we filter their games only. We do not select matches where one of the players withdrawn or did not complete the match because of an injury, therefore only completed matches are taken into account. We also removed matches from Davis Cup and the Olympic Games. Moreover, because one of the variables of interest is the surface and since the number of games on carpet was very small for these players, since they ATP stopped using this material in 2009, we also removed matches played on that surface. After this selection process, we have a total of 3,653 observations.

We are specially interested in variables regarding minutes played and relative points won in each match. Although minutes could be directly affected by the differences in how each player prepares for each point played, as far as how much time they take in between points, but also in their approach throughout the match, we still consider that this information could be connected to how it was easily won by one of the players (not necessarily Djokovic, Federer or Nadal) or how hard the match was. In this case, less minutes in court could be linked to a greater dominance, while more minutes could be associated with less dominance or a closer match. Nevertheless, this needs to be controlled by the surface, since clay courts games usually take longer to finish than those played on grass or hard courts, for instance. Moreover, grand slam tournaments are played as best of 5 sets, instead of best of 3, which is the case for most of the tournaments. Therefore, we need to take into account the tournament type as well.

\begin{figure}
\centering
\includegraphics[scale=0.5]{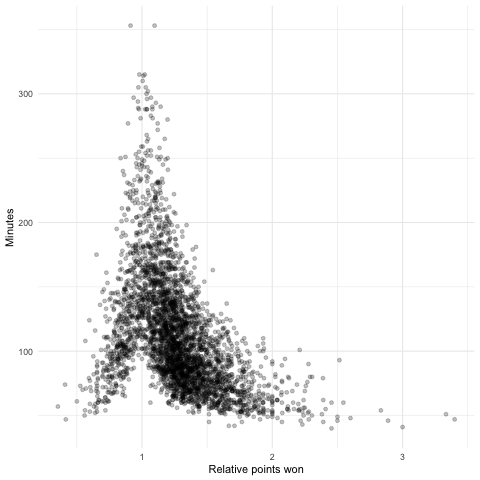}
\caption{\label{dispersion_plot} Scatter plot between relative points won and minutes played for all completed matches for Djokovic, Federer and Nadal.}
\end{figure}

In Figure~\ref{dispersion_plot}, we can see the scatter plot between these two variables. One is able to see that longest matches are closer to the value 1.0 of relative points won. In fact, the longest match in the data set is related to the Australian Open final between Djokovic and Nadal in 2012, where they battled for almost 6 hours. In this case, this match is duplicated at the highest value for the $y$ axis, where the difference in $x$ axis shows the two values for relative points for each player; by definition, these values are symmetrical around 1. For this particular match, Djokovic won the match and also around 9\% more points than Nadal, which puts relative points won close to the value 1.09.

We can also note that matches, where these players had a bigger (or smaller) value for relative points won, had a lower minutes value, which seems appropriate. If one of the players wins (or loses) the majority of the points, this match should finish in fewer minutes. Though this is certainly not a sufficient condition for faster matches. It is important to emphasise that points themselves are not exactly connected to winning or losing, since players need to win games, in order to win sets, in order to win the match. There is surely a connection between these outcomes, but this is not exact. For instance, for these three great players and their completed matches shown in Figure~\ref{dispersion_plot}, in around 10\% of their losses they still managed to win more points than their opponent throughout the match. This number is around 5\% if we consider all matches in this period in the ATP tour. Due to this higher performance even in losses, in comparison with the rest of the ATP tour, we believe it is worth considering as one of our covariates, whether the match was won or lost by the big three. Because tennis is a sport where a loss is equivalent to an elimination of the tournament, with the exception of tournaments like Finals, it is intriguing to compare their performance in these different types of games.

Given the different types of numbers of players involved and even participation of top players, we split the tournaments in 4 different types: Grand Slam, Masters, Finals and others. Grand Slams are the most important and the most prestigious tournaments and they are played as a best of five sets format. Therefore, it is important that they are on their own category. Masters tournaments are currently known as Masters 1000 and there are nine tournaments of this level throughout the year, being played in all different surfaces and continents around the world. Finals is just one tournament played with the best 8 players of each year. They are played in a different structure, in a round robin format, where players are divided in two groups and the best two of each group are headed to the semifinals, after all playing against each other inside the group. Besides these tournaments, we consider all other tournaments in the same category.
§
If we want to discuss dominance against other players, it is important to try to measure how each one of these three great tennis players fared against different type of competition. Besides considering the effects for surface and types of tournaments, we split their matches against top 20 players in the world rankings and other players. This is to show how they contrast when facing a more challenging opponent, given by a higher ranking, in comparison with games against players not so well ranked. This is not an admission of easier games against opponents outside this interval, but rather a remark about the high quality of play that tennis players must attain to reach a top 20 world ranking. When those players reach this level, one anticipates tight matches and it is important to compare the performance of the big three against this level of competition.

We could have used age as well as one of our covariates, but we decided to omit that information given the rather amazing career path that each one of these players took while ageing in a different manner than most players. For instance, \citet{gorgi:19} estimates that players on average reach their highest performance around age of 25 years. While this is possibly true for most of the players, it might not hold true to the big three. Federer and Nadal first achieved the number one rankings at the age of 22 years old, while Djokovic reached the top at 24 years old. Between 2004 and 2020, they have been swapping this first position continuously, except for one year, when Andy Murray finished the year as the best player in the planet. They are now the oldest players ever to hold the number one ranking, with Federer currently leading this category when he obtained the highest position in the ranking at the age of 36 years old. Given these circumstances, we decide to focus their comparisons taking into consideration only other variables described here in this section. 

\section{Bayesian quantile regression for multiple output response variables} \label{sec_theory}

Considering our two dimensional response variable, relative points won and minutes, we would like to describe their joint variation as a function of all the variables described in the previous section. Given their variation presented in Figure~\ref{dispersion_plot}, we would like to study their conditional probability distribution, without assuming any probability distribution at first. Taking that into consideration, we consider a Bayesian form of the directional quantile regression model for multiple outputs proposed by \citet{hallin:10}. We use the same approach proposed by \citet{santos:20}, where an asymmetric Laplace distribution is considered in each direction of interest in the likelihood of our model.

Regarding this model, we need to define a few quantities. First, let a directional index be ${\boldsymbol \tau} \in \mathcal B^k := \{ {\boldsymbol v} \in \mathbb{R}^k: 0 < || {\boldsymbol v} ||_2 < 1 \}$, a collection of vectors encompassed in the unit ball of $\mathbb{R}^k$. We can split this directional index in two parts, ${\boldsymbol \tau} = \tau {\boldsymbol u}$, where ${\boldsymbol u} \in \mathcal{S}^{k-1} :=  \{\boldsymbol z \in \mathbb{R}^k : ||\boldsymbol z|| = 1 \}$, represents the direction and $\tau \in (0,1)$ the magnitude of interest. Moreover, let let $\boldsymbol \Gamma_u$ be an arbitrary $k \times (k-1)$ matrix of unit vectors, where 
$(\boldsymbol u \, \vdots \, \boldsymbol \Gamma_u)$ establishes an orthonormal basis of $\mathbb{R}^k$. Define ${\boldsymbol Y_u} := {\boldsymbol u}^{'} {\boldsymbol Y}$ and ${\boldsymbol Y}^{\perp} := {\boldsymbol \Gamma_u}^{'} \boldsymbol Y$. 

Therefore, for each direction we consider a mixture representation of the asymmetric Laplace proposed by \citet{kozumi:11}, as the following 
\begin{align*} \label{transformedModel}
 Y_u | \boldsymbol b_\tau, \boldsymbol \beta_\tau, \sigma, \boldsymbol v &\sim N(\boldsymbol \eta_u + \theta \boldsymbol v, \psi^2 \sigma \boldsymbol V), \\
 v_i &\sim \mbox{Exp}(\sigma), \quad i=1,\ldots,n, 
\end{align*}
where $ \theta = (1-2\tau)/(\tau(1-\tau))$, $\psi^2 = 2/(\tau(1-\tau))$, $\mbox{Exp}(\sigma)$ denotes the exponential distribution with mean $\sigma$, $V = \mbox{diag}(v_1, \ldots, v_n) $ and $\boldsymbol \beta_\tau$ contains a intercept. Our linear predictor for this model is defined as
\begin{equation*}
 \boldsymbol \eta_u = X \boldsymbol \beta_\tau + Y_u^\perp \boldsymbol b_\tau,
\end{equation*}
where $X$ is a $n \times p$ design matrix containing our covariates and an intercept.

Then one needs to define priors distributions for $\xi = (\boldsymbol b_\tau, \boldsymbol \beta_\tau, \sigma)$ to complete the specification of the model. For $\boldsymbol b_\tau$, \citet{guggisberg:19} discusses its prior elicitation, where the author relates this prior to the Tukey depth of the data. Combined with $\boldsymbol \beta_\tau$, we use a multivariate normal distribution with mean 0 and a large variance for these parameters for $(\boldsymbol b_\tau, \boldsymbol \beta_\tau)$. For $\sigma$, we consider an inverse gamma distribution as in \citet{waldmann2013}. Inference on the parameters is based on samples of the posterior distribution obtained via Gibbs sampling and we refer to these previous references for more details. Our estimates for each parameter will be defined as the posterior means, namely $(\hat{\boldsymbol b}_\tau, \hat{\boldsymbol \beta}_\tau, \hat{\sigma})$. 

Given these parameters estimates, which are obtained for each direction $\boldsymbol u$, we are able to define an upper closed quantile halfspace
\begin{equation} \label{halfspace}
 H^+_{\tau \boldsymbol u} = H^+_{\tau \boldsymbol u} (\hat{\boldsymbol b}_\tau, \hat{\boldsymbol \beta}_\tau) = \{ \boldsymbol y \in \mathbb{R}^k : \boldsymbol u^{'} \boldsymbol y \geq \hat{\boldsymbol b}_\tau \boldsymbol \Gamma^{'}_u \boldsymbol y + \boldsymbol x^{'} \hat{\boldsymbol \beta}_\tau \}
\end{equation}
and a lower open quantile halfspace switching $\geq$ for $<$ in \eqref{halfspace}. Moreover, for fixed $\tau$ we define the $\tau$ quantile region $R(\tau)$ as 
\begin{equation} \label{tauRegion}
 R(\tau) = \bigcap_{{\boldsymbol u} \in \mathcal{S}^{k-1}} H_{\tau \boldsymbol u}^+.
\end{equation}

These quantile regions are important as we use them to study the effect of each predictor variable in the conditional distribution of $\boldsymbol Y$. \citet{hallin:10} showed that this quantile region are equal to Tukey depth regions, which have been extensively studied to describe the probability distribution of multivariate distributions. In our case, we consider the boundary of these regions, also called quantile contours to make the comparison between the results obtained by the three great players. Even though \citet{hallin2017} alert that severe assumptions are needed in order for these quantities to be the depth contours of the conditional distribution of $\boldsymbol Y$ given $\boldsymbol X$, we follow \citet{santos:20} and consider that these can be considered an averaged version of the Tukey depth contour. We consider the same algorithm proposed by the latter and in the next section we will use these quantile contours to make the comparisons between Djokovic, Federer and Nadal, conditional on the variables described in Section~\ref{data_section}.

\section{Results}

For our application, we estimate Bayesian quantile regression models for multiple output as described in the previous section and show here the quantile contours for different combinations of the covariates. We consider the data outlined in Section~\ref{data_section}, where our main interest lies in the bivariate response variable, relative points won and minutes. 

Four our modelling purposes, we consider 180 directions defined uniformly in the unit ball. We take $\tau = 0.25$, so we can have a more central quantile region to compare their results. We have tried other values for $\tau$ and the conclusions are quite similar. For each direction, for each chain we consider a burn-in size of 10,000 and we run the chain to obtain 100,000 observations, keeping every 100th draw from this posterior.

We consider the following covariates for our model, with reference category denoted inside the parenthesis. We use main effects to denote players (Federer - reference, Djokovic, Nadal); win (no - reference, yes); surface (hard - reference, clay, grass); tournament (others - reference, Grand Slam, Finals, Masters); top 20 (no - reference, yes). Given our objective of comparing the three players, we add an interaction term between player and all other covariates. Although we do not the equation here, the idea is easy to follow, as we need to have a dummy variable for each one of the categories that are not defined as the reference. Besides, there needs to be a coefficient for each one of the interactions. In total, for this regression part of the model we are interested in 24 parameters plus the coefficient related to the direction projection $Y_u^\perp$. Therefore, to avoid a long equation we decided to skip its definition here. We have to denote also that each one of these parameters is attached to a direction and the visualisation (or the analysis) is possible via the comparison of the quantile regions.

For each direction, we considered a transformation of the response variable to make computations more stable. Instead of using each variable in their original scale, we stardardized their value, i.e., we centered each variable by subtracting its mean and divided by its standard deviation. In order to show the obtained quantile contours, we still consider this range of variation, though one could also transform to the original scale given how quantile are equivariant to monotone transformations. Because we are considering the transformed values, we must always compare the variation among the different quantile contours for the different players, instead of their actual values. Moreover, in order to observe the effects of each covariate, we plot its respective quantile contours for the three players, but we keep the values for all the other covariates at their reference value. For instance, the reference value for variable win is no, then for this reason the range of values in the $x$ axis will be mostly negative. Likewise, we reiterate that the differences among the different players and the different covariate values is more important than the value itself.

\begin{figure}
\centering
\includegraphics[scale=0.5]{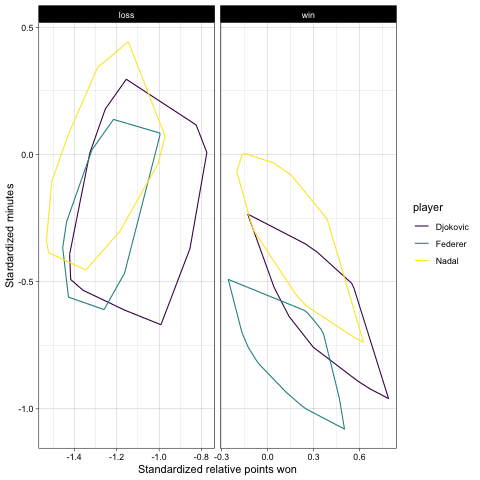}
\caption{\label{effect_win} Quantile contours to compare the performance between Djokovic, Federer and Nadal in their losses and wins. For all other variables we consider their reference value.}
\end{figure}

We move to our conclusions first considering the quantiles contours that compare their performances in wins and losses presented in Figure~\ref{effect_win}. In the dimension of time in wins, there is a clearly advantage for Federer, as his quantile contours presents the smallest values for that dimension. For the other variable, relative points won, for either losses or wins, Djokovic has an advantage over the other two. When we compare the quantile contours of Federer and Nadal, they have a quite similar structure, but Nadal's contour is moved upwards in the time dimension. In wins though, Nadal shows a small advantage over Federer in the relative points won dimension. Another interesting point in this comparison is how the correlation between these two variables seems to be higher for wins rather than losses. In this case, for wins it is easier to see that as the relative points won increases, the minutes variable decreases, which agrees with a common intuition about the game: when these three win most of the points the match will end more rapidly. The same cannot be said about losses. 

\begin{figure}
\centering
\includegraphics[scale=0.5]{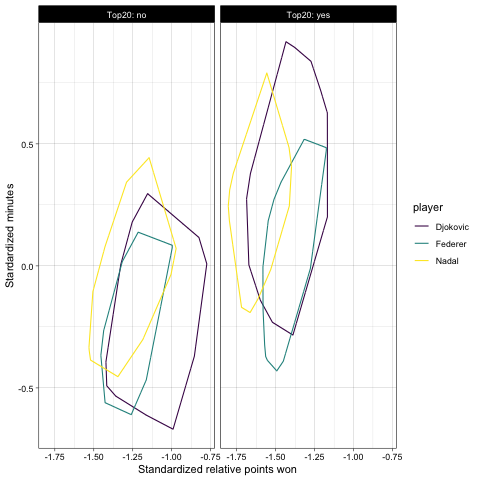}
\caption{\label{effect_top20} Quantile contours to compare the performance between Djokovic, Federer and Nadal in their matches against players in top 20 positions of the world rankings or not. For all other variables we consider their reference value.}
\end{figure}

When we consider the effect of playing against top 20 players in the ranking or not, we can compare the quantile contours presented in Figure~\ref{effect_top20}. A first look shows that in the time dimensions for all three players there is a shift upwards when we compare their performance against the best players in comparison with not so high ranked players. Moreover, in the relative points won dimension, there is a shift to the left, meaning that these games are possibly more difficult, or at least closer in this sense, against high ranked players. Again, this would follow our common intuition about their results. Furthermore, still in the relative points won perspective, while Djokovic seems to have an advantage against lower ranked players, Federer presents a quantile region with less variance and with the highest values equivalent to Djokovic highest values.

\begin{figure}
\centering
\includegraphics[scale=0.5]{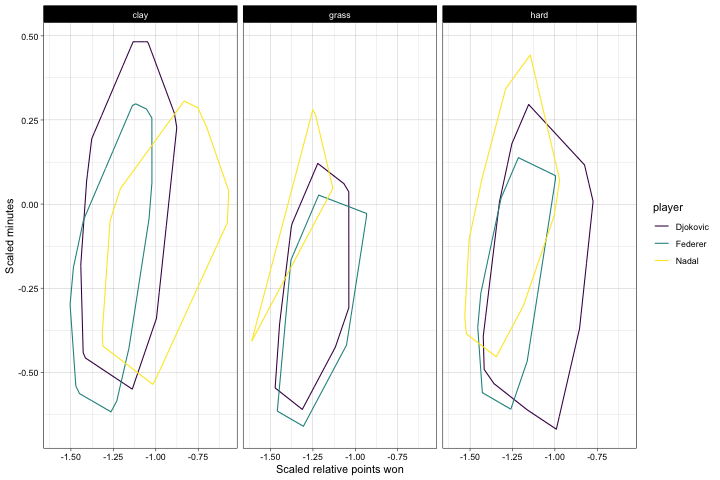}
\caption{\label{effect_surface} Quantile contours to compare the performance between Djokovic, Federer and Nadal in their matches considering the three different surfaces. For all other variables we consider their reference value.}
\end{figure}

In the comparison about surfaces, which is shown in Figure~\ref{effect_surface}, we have an interesting expression of the dominance of each player in a different surface. First, only analysing the relative points won dimension, Nadal is far superior in clay, Federer leads in grass courts and Djokovic is ahead in hard courts. This is expected since Nadal is the possibly the best ever player in clay courts, where he has 13 French Open titles, the Grand Slam played in that surface. Meanwhile, Federer has the most titles in Wimbledon, which is played in grass courts and Djokovic leads the titles in the Australian Open, a hard court event. Both Federer and Djokovic have a good number of titles in the US Open, another event in hard court, but that does not make their difference smaller. In fact, Nadal is slight ahead of Federer in that direction. Regarding minutes spent on court, Federer shows smaller values than the other in clay and grass court games. For games played on hard courts, Djokovic does attain the smallest of the values in this dimension on hard courts, though Federer's quantile region is contained in a smaller space, which indicates a smaller variation of values. Interestingly, when one compares the different quantile regions across the surfaces, Nadal's case in clay courts, seems to be the most far off from the other. His quantile region is moved completely to the right in comparison with the other. Djokovic dominance in hard courts is quite big too, but for some directions his quantile region is not too distant from Federer region. This is an evidence to describe that even though each one has an edge on a particular surface, Nadal's clay court dominance possibly is the biggest of them all.

\begin{figure}[!htb]
\centering
\includegraphics[scale=0.5]{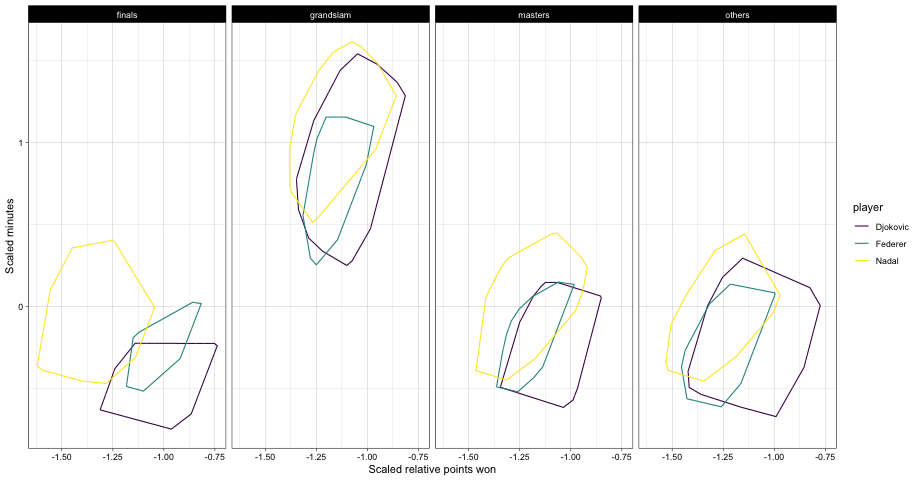}
\caption{\label{effect_tournament} Quantile contours to compare the performance between Djokovic, Federer and Nadal in their matches considering four different types of tournaments. For all other variables we consider their reference value.}
\end{figure}

Regarding the different tournament types, one can easily perceive the difference between Grand Slam tournaments and the other tournaments in the minutes dimension, displayed in Figure~\ref{effect_tournament}. All quantile regions from Grand Slam tournaments for the three players are noticeably above the other tournaments. This is no surprise given the best of five sets format for the more prestigious tournaments. Still on the time dimension, Nadal appears to be the player with higher minutes on court, while Djokovic is the one who gets the smallest values in his region. In the relative points won dimension, Djokovic shows the highest values for all types of tournaments, with Nadal in second in all types except the Finals. Notably, Nadal quantile region for this tournament is rather different than the other players. In fact, Nadal has no titles in this type of tournament, which could explain this result. Overall, one would give Djokovic the acknowledment as the most dominant in all types of tournaments, when controlling with the other variables.

\section{Concluding remarks} \label{concluding}

For this manuscript, we made an effort to discuss which one of the these three great players of tennis, Djokovic, Federer and Nadal, is more dominant. We used the information about their minutes and relative points won in each one of their matches and considered some covariates, such as win or losses, surfaces, top 20 opponent or not and type of tournament. By considering a Bayesian regression model for multiple output response variables we were able to compare their results jointly regarding these two response variables, without assuming a joint probability distribution for our response variable. This presents a more flexible way of comparing this conditional probability distributions of interest with less restrictions about their exact shape, for instance. Conclusions about the results of each player were based on the comparison of their quantile region conditional on the chosen covariates.

There are a few interesting results, which are worth mentioning given the plots presented in the previous section. While each player showed some dominance in one specific surface, Nadal's superiority in clay court is simply above the others. His unprecedented record of 13 French Open titles and many other titles in clay courts probably does support this result. Federer quickness in wins, with his quantile region shifted downwards in comparison with the other two, is certainly meaningful. This could be attributed perhaps to his playing style, since Nadal and Djokovic are known to rely more heavily in their remarkable defensive skills. With the results presented here, Djokovic should be considered the most dominant of the three. Specially, when one considers the relative points won dimension, which seems to be a more powerful quantity to measure this information. For this dimension, Djokovic is ahead in many of the combination of covariates considered. This is possibly a testament to the several accolades he has received during his accomplished career. He is, currently, the only one between the three who has won each Grand Slam tournament at least twice and also the only who won at least once all the Masters tournaments. This is certainly one way of defining dominance.

This is possibly not the last comparison of these great players, but this application was able to show some fascinating results for these great players. Though the measures used here to describe dominance have their weaknesses, we believe they still give an interesting approximation of superiority against their opponents. We hope that this manuscript, where we used a different approach, not only regarding the modelling technique, but also the chosen variables, leads to other stimulating discussions in this area.


\begin{thebibliography}{14}
\providecommand{\natexlab}[1]{#1}
\providecommand{\url}[1]{\texttt{#1}}
\providecommand{\urlprefix}{URL }
\providecommand{\selectlanguage}[1]{\relax}

\bibitem[\protect\citename{Baker and McHale, }2014]{baker:14}
Baker, R.D., McHale, I.G.:  (2014), A dynamic paired comparisons model: Who is
  the greatest tennis player? \emph{European Journal of Operational Research}
  \textbf{236}, 677--684.

\bibitem[\protect\citename{{del Corral} and Prieto-Rodriguez,
  }2010]{delCorral:10}
{del Corral}, J., Prieto-Rodriguez, J.:  (2010), Are differences in ranks good
  predictors for grand slam tennis matches? \emph{International Journal of
  Forecasting} \textbf{26}, 551--563, sports Forecasting.

\bibitem[\protect\citename{Gorgi et~al., }2019]{gorgi:19}
Gorgi, P., Koopman, S.J., Lit, R.:  (2019), The analysis and forecasting of
  tennis matches by using a high dimensional dynamic model. \emph{Journal of
  the Royal Statistical Society: Series A (Statistics in Society)}
  \textbf{182}, 1393--1409.

\bibitem[\protect\citename{{Guggisberg}, }2019]{guggisberg:19}
{Guggisberg}, M.:  (2019), {A Bayesian Approach to Multiple-Output Quantile
  Regression}. \emph{arXiv e-prints} arXiv:1909.02623.

\bibitem[\protect\citename{Hallin et~al., }2010]{hallin:10}
Hallin, M., Paindaveine, D., {\v S}iman, M.:  (2010), Multivariate quantiles
  and multiple-output regression quantiles: From l1 optimization to halfspace
  depth. \emph{The Annals of Statistics} \textbf{38}, 635--669.

\bibitem[\protect\citename{Hallin and {\v S}iman, }2017]{hallin2017}
Hallin, M., {\v S}iman, M.:  (2017), Multiple-output quantile regression.
  \emph{In:} \emph{Handbook of quantile regression}, edited by R.~Koenker,
  V.~Chernozhukov, X.~He, L.~Peng, Chapman and Hall/CRC.

\bibitem[\protect\citename{Klaassen and Magnus, }2003]{klaassen:03}
Klaassen, F.J., Magnus, J.R.:  (2003), Forecasting the winner of a tennis
  match. \emph{European Journal of Operational Research} \textbf{148},
  257--267.

\bibitem[\protect\citename{Klaassen and Magnus, }2001]{klaassen:01}
Klaassen, F.J.G.M., Magnus, J.R.:  (2001), Are points in tennis independent and
  identically distributed? evidence from a dynamic binary panel data model.
  \emph{Journal of the American Statistical Association} \textbf{96}, 500--509.

\bibitem[\protect\citename{Knottenbelt et~al., }2012]{knottenbelt:12}
Knottenbelt, W.J., Spanias, D., Madurska, A.M.:  (2012), A common-opponent
  stochastic model for predicting the outcome of professional tennis matches.
  \emph{Computers \& Mathematics with Applications} \textbf{64}, 3820--3827.

\bibitem[\protect\citename{Koenker and Bassett, }1978]{koenker:78}
Koenker, R., Bassett, G.:  (1978), Regression quantiles. \emph{Econometrica}
  \textbf{46}, 33--50.

\bibitem[\protect\citename{Kozumi and Kobayashi, }2011]{kozumi:11}
Kozumi, H., Kobayashi, G.:  (2011), Gibbs sampling methods for {B}ayesian
  quantile regression. \emph{Journal of Statistical Computation and Simulation}
  \textbf{81}, 1565--1578.

\bibitem[\protect\citename{McHale and Morton, }2011]{mchale:11}
McHale, I., Morton, A.:  (2011), A {B}adley-{T}erry type model for forecasting
  tennis match results. \emph{International Journal of Forecasting}
  \textbf{27}, 619--630.

\bibitem[\protect\citename{Santos and Kneib, }2020]{santos:20}
Santos, B., Kneib, T.:  (2020), Noncrossing structured additive multiple-output
  {B}ayesian quantile regression models. \emph{Statistics and Computing}
  \textbf{97}, 825.

\bibitem[\protect\citename{Waldmann et~al., }2013]{waldmann2013}
Waldmann, E., Kneib, T., Yue, Y.R., Lang, S., Flexeder, C.:  (2013), Bayesian
  semiparametric additive quantile regression. \emph{Statistical Modelling}
  \textbf{13}, 223--252.

\end{thebibliography}
\end{document}